\documentclass[showpacs,preprintnumbers,amsmath,amssymb,twocolumn,aps,pre]{revtex4-1}
\usepackage{color}
\usepackage{amsmath}
\usepackage{mathrsfs}
\usepackage{graphicx}
\usepackage{subfig}
\usepackage{epsfig}

\begin{document}
\title{Formation of localized states in dryland vegetation: Bifurcation structure and stability}
\author{P. Parra-Rivas$^{1,2}$, C. Fernandez-Oto$^{3}$}

\affiliation{
	$^1$Service OPERA-photonics, Université libre de Bruxelles, 50 Avenue F. D. Roosevelt, CP 194/5, B-1050 Bruxelles, Belgium\\
	$^2$Laboratory of Dynamics in Biological Systems, Department of Cellular and Molecular Medicine, University of Leuven, B-3000 Leuven, Belgium\\
	$^3$ Complex Systems Group, Facultad de Ingenieria y Ciencias Aplicadas, Universidad de los Andes, Av. Mon. Alvaro del Portillo 12455 Santiago, Chile}
\date{\today}


\begin{abstract}
In this paper, we study theoretically the emergence of localized states of vegetation close to the onset of desertification. These states are 
formed through the locking of vegetation fronts, connecting a uniform vegetation state with a bare soil state, which occurs nearby the Maxwell point of the system. To study these structures we consider a universal model of vegetation dynamics in drylands, which has been obtained as the normal form for different vegetation models. Close to the Maxwell point localized gaps and spots of vegetation exist and undergo collapsed snaking. The presence of gaps strongly suggest that the ecosystem may undergo a recovering process. In contrast, the presence of spots may indicate that the ecosystem is close to desertification.
\end{abstract}
\maketitle

\section{Introduction}

Localized structures, hereafter LSs, are a particular type of the more general dissipative states that emerge naturally in extended systems far from thermodynamic equilibrium \cite{nicolis_self-organization_1977,cross_pattern_1993}.
LSs are not related with the intrinsic inhomogeneities of the system, but arise due to a double balance between nonlinearity and spatial coupling (e.g. diffusion) on one hand, and energy gain and dissipation on the other hand \cite{akhmediev_dissipative_2005}. Their formation is usually related to the presence of bi-stability between two different stable states. In this context, a LS can be seen as a localized portion of a given domain embedded within a different one.

Examples of LSs can be found in a large variety of natural systems ranging from optics and material science to population dynamics and ecology \cite{akhmediev_dissipative_2005,dawes_j._h._p._emergence_2010,knobloch_spatial_2015}.
In plant ecology, LSs have been observed in different contexts such as drylands \cite{macfadyen_vegetation_1950,becker_fairy_2000,van_rooyen_mysterious_2004,meron_localized_2007,VegBxl2011Deblauwe,meron_pattern-formation_2012,tschinkel_experiments_2015,getzin_discovery_2016} and marina sea-grass \cite{ruiz-reynes_fairy_2017} ecosystems. In particular, in arid and semi-arid regions, LSs can appear as spots \cite{lejeune_localized_2002, escaff_localized_2015, zelnik_yuval_r._regime_2013}, gaps \cite{tlidi_vegetation_2008, Veg_FC_2014_Fernandez-Oto, zelnik_localized_2016}, rings \cite{VegIsrael2007Sheffer2007, VegIsrael2011Sheffer,VegIsrael2019YizhaqRings}, and spirals \cite{Fernandez-Oto2019Spirals} among others. 

These ecosystems are exposed to desertification processes which can take place through the slow advance of the barren state, i.e. front propagation \cite{zelnik_desertification_2017} or abrupt collapse \cite{VegOtros2012Scheffer,VegIsrael2012Bel}, and therefore their study is very relevant. 

One potential scenario for the formation of LSs is related to the presence of bi-stability between a uniform, and a spatially periodic (pattern) state emerging from a Turing instability \cite{turing_alan_mathison_chemical_1952}. In the context of semi-arid ecosystems,
spatially periodic patterns have been widely studied \cite{macfadyen_vegetation_1950,VegBxl1997Lefever,VegOtros1999Klausmeier,VegRietkerk2002Rietkerk,VegRietkerk2004Rietkerk,VegStochastic2006DOdoricoJournal,VegRietkerk2008Rietkerk,VegStochastic2009Borgogno,VegBxl2011Deblauwe,VegIsrael2014Yizhaq,VegBxl2014Couteron,VegPalma2014Martinez-Garcia,VegPalma2013Martinez-Garcia_Calabrese1,VegOtros2014Gowda,VegIsrael2015meron}, and the formation of LSs in this context has been investigated \cite{lejeune_localized_2002,tlidi_vegetation_2008, zelnik_yuval_r._regime_2013, zelnik_localized_2016}. 
Furthermore, the formation of LSs has been also analyzed in the presence of strong nonlocal coupling \cite{Veg_FC_2014_Fernandez-Oto,escaff_localized_2015}.

Another plausible bi-stable scenario for the formation of LSs is based on the locking of fronts that connect two different, but coexisting, uniform states. This mechanism is well understood and  
has been widely studied in different physical and natural systems \cite{coullet_nature_1987,oppo_domain_1999,oppo_characterization_2001, coullet_localized_2002,clerc_patterns_2005,escaff_non-local_2011,colet_formation_2014, parra-rivas_dark_2016}:
near the Maxwell point of the system \cite{Goldstein1991pra,Fernandez-Oto2013prl}, and in the presence of spatial damped oscillations around the uniform states, fronts interact and lock at different separations, 
leading to the formation of LSs.
In semi-arid ecosystems, the behavior of these vegetation fronts on either sides of the Maxwell point is closely related to the beginning of a gradual desertification process, and therefore its understanding is of great importance. Despite this relevance, only a few works focus on the formation of such types of states in dryland ecosystems, in particular in the Gray-Scott model \cite{gandhi_punit_spatially_2018,zelnik_implications_2018}. 
Here we present a detailed study of these LSs focusing on their origin, bifurcation structure, and stability.

The article is organized as follows. In Sec.~\ref{sec:1} we introduce the model. 
Section~\ref{sec:2} is devoted to the study of homogeneous or uniform solutions of the system, their spatio-temporal stability, and their spatial dynamics. 
Later, in Sec.~\ref{sec:3} we introduce the mechanism of front locking to describe the formation of LSs. 
In Sec.~\ref{sec:4}, applying multi-scale perturbation theory, we calculate small amplitude weakly non-linear solutions near the main bifurcations of the uniform state.
Following on from this, in Sec.~\ref{sec:5} and \ref{sec:6} we build the bifurcation diagrams of the different LSs found analytically, and classify their regions of existence in the parameter space. 
Finally, in Sec.~\ref{sec:7} we present a short discussion and the conclusions.

\section{The model}\label{sec:1}
Dryland vegetation ecosystems are a particular type of pattern-forming living systems. One characteristic of these systems is that the typical state variables, such as population densities of organisms and biochemical reagents concentrations, cannot assume negative values.
\\

In this context we study the reduced model
\begin{equation}\label{normal}
\partial_tA=\alpha A+\beta A^2-A^3+D\nabla^2 A-A\nabla^2A-A\nabla^4A,
\end{equation}
where $\nabla^2\equiv\partial_x^2+\partial_y^2$, $A$ is a $\mathbb{R}^+$-valued scalar field, and $\alpha,\beta$, and $D$ are the three real control parameters of the system. 
In general this model can be derived close to the onset of bi-stability where nonviable states undergo subcritical instabilities to viable states.

In the context of dryland vegetation, Eq.~(\ref{normal}) was initially derived from the Lefever-Lejeune model \cite{VegBxl1997Lefever} in the weak-gradient approximation \cite{lejeune_model_1999,tlidi_vegetation_2008}, and just recently from the Gilad model \cite{gilad_ecosystem_2004, VegIsrael2007Gilad,fernandez-oto_front_2019}.
Furthermore, similar models have been found in other living systems with Lotka-Volterra type of dynamics \cite{paulau_self-localized_2014} and in sea-grass marina ecosystems \cite{ruiz-reynes_simple_2019}.
Note that a similar equation has been also derived in non-living systems \cite{kozyreff_nonvariational_2007}.

 	

In this work we analyze Eq.~(\ref{normal}) in the framework of dryland vegetation, and we refer to the derivation obtained in \cite{fernandez-oto_front_2019}. In this context, $A$ is proportional to the biomass density, and is a positive quantity; $D>0$ is proportional to the ratio between the biomass diffusion (lateral growth and seed dispersion) and the soil water diffusion; $\beta>0$ quantifies the subcriticality of the uniform vegetation state and is related to the root-to-shoot ratio; and $\alpha$ measures the distance to the critical precipitation point where the bare state changes its stability.

Due to the complexity of this model, we focus on the study of Eq.~(\ref{normal}) in one spatial dimension, and hereafter we consider $\nabla^2\equiv\partial_x^2$. In what follows we consider finite, although very large, domains with periodic boundary conditions.

\section{The homogeneous solutions and their linear stability}\label{sec:2}
In this section we first introduce the homogeneous solutions of the system, we perform a spatio-temporal stability analysis, and we study their spatial dynamics. In this way we are able to identify the different stability regions, and the bifurcations from where small amplitude LSs may potentially arise.
\begin{figure}[t]
	\centering
	\includegraphics[scale=1]{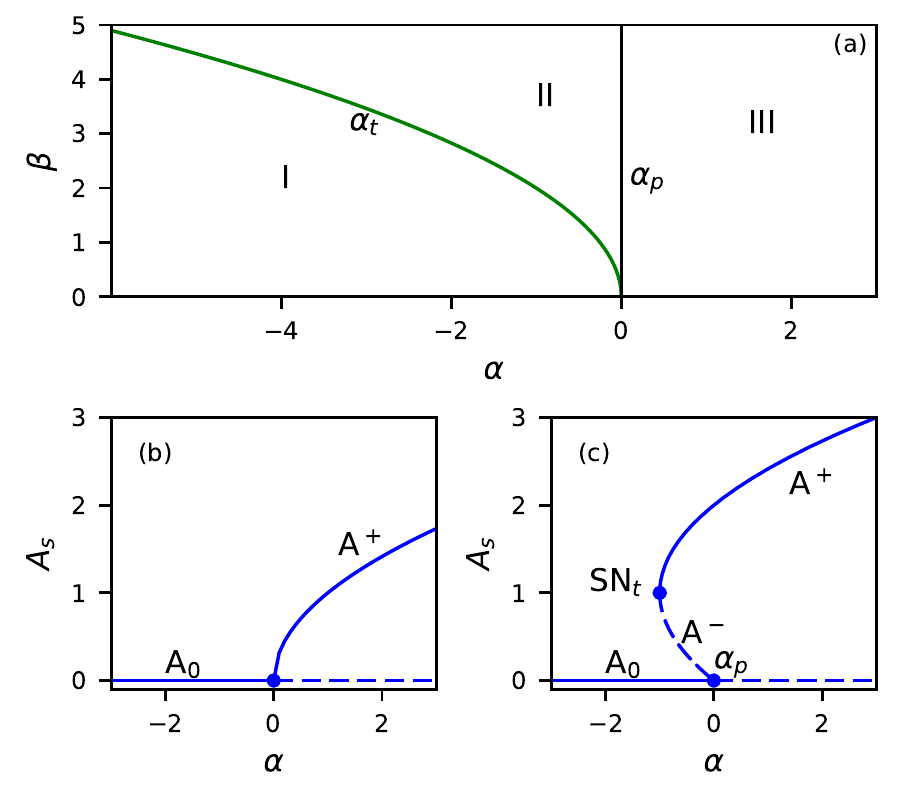}
	\caption{(Color online) In (a) the phase diagram in the $(\alpha,\beta)-$parameter space showing the principal bifurcation lines of the HSS solutions: the pitchfork bifurcation $\alpha_p$ (black line), and the fold line $\alpha_t$ corresponding to SN$_t$ (green line). (b) shows the HSS solutions for $\beta=0$, and (c) those for $\beta=2$. The linear stability respect to homogeneous perturbations is shown using solid (dashed) lines for stable (unstable) states. The different regions are labeled by I-III and their description is given in the main text.}
	\label{F_sol_hom}
\end{figure}

\subsection{Homogeneous steady states}

In this subsection we consider the system without space, and we calculate their homogeneous states solutions and their stability. 
The homogeneous steady state (HSS) solutions of the system satisfy 
\begin{equation}
-A_s(A_s^2-\beta A_s-\alpha)=0,
\end{equation}
and therefore consist of three branches of solutions: the trivial state $A_s=A_0\equiv0$, representing {\it bare soil}, and the two non-trivial state branches
\begin{equation}
A^{\pm}=\frac{\beta\pm\sqrt{\beta^2+4\alpha}}{2},
\end{equation}
which represent {\it uniform vegetation} states. Note that, any solution $A_s$ should be zero or positive because negative biomass does not exist. It is easy to prove that $A_0=0$ is stable (unstable) when $\alpha<0$ ($\alpha>0$). It is also straightforward to obtain that $A^+$ is always stable, and $A^-$ is always unstable, when they are positive.

The branches $A^+$ and $A^-$ are separated by a fold or saddle-node (SN$_t$) point occurring at
\begin{equation}\label{position_SN}
A_t=\frac{\beta}{2}, \hspace{1.1cm} \alpha_t=\frac{-\beta^2}{4}.
\end{equation}
Furthermore, the system undergoes a transcritical bifurcation at $\alpha=\alpha_p\equiv0$. 

In the $(\alpha,\beta)-$parameter space these bifurcations define the two lines shown in the phase diagram of Fig.~\ref{F_sol_hom}(a).  Slices of constant $\beta$ correspond to the bifurcation diagrams shown in 
Fig.~\ref{F_sol_hom}(b) and Fig.~\ref{F_sol_hom}(c), for $\beta=0$, and $\beta=2$ respectively. The point where the system changes from a subcritical to a supercritical bifurcation corresponds to the value $\beta=0$. 

At this stage we can identify three different regions that we label as I-III:
\begin{itemize}
	\item Region I: Only the bare soil $A_0$ exists and is stable. This region is spanned by parameter values below the fold line $\alpha_t$ when $\beta>0$, and  by $\alpha_p$ when $\beta=0$.
	\item Region II: The non-trivial states $A^+>0$ and $A^->0$ coexist with $A_0$. This region is spanned by $\alpha_t<\alpha<\alpha_p$ and $\beta>0$. This is the bi-stability region.  The solutions $A_0$ and $A^+$ are stable.
	\item Region III: Only solutions $A_0$ and $A^+$ coexist, but now $A_0$ is always unstable. This region is defined by $\alpha>\alpha_p$. 	
\end{itemize}
In this work we are interested in the region of parameters where fronts connecting the bare soil and the homogeneous vegetation exist, i.e. the bi-stability region II. As we will see in the following sections, desertification and recovery fronts exist in this region, and their interaction is related with the formation of LSs. 
\subsection{Spatiotemporal stability analysis}
The next step in our study is to analyze the linear stability of the HSS solutions $A_s$ against general non-uniform perturbations $\xi=e^{\sigma t+ikx}$, or what is the same, to consider weakly modulated solutions of the form  $A(x,t)=A_s+\epsilon \xi(x,t)$, with $|\epsilon|\ll1$. Considering this ansatz the stability problem reduces to the study of the linear problem 
\begin{equation}
\partial_t\xi=\mathcal{L}\xi,	
\end{equation}
where $\mathcal{L}$ is the linear operator 
	\begin{equation}\label{lin_op}
	\mathcal{L}\equiv \alpha+ 2\beta A_s-3A_s^2+D\partial_x^2-A_s\partial_x^2-A_s\partial_x^4.
	\end{equation}
For this equation one obtains that the growth rate $\sigma$ satisfies
\begin{equation}\label{sigma}
\sigma(k)=\alpha+2\beta A_s-3 A_s^2-D k^2+A_s k^2-A_s k^4.
\end{equation}
The HSS $A_s$ is stable when ${\rm Re}[\sigma(k)]<0$, and unstable otherwise. For a finite value of $k$, the transition between these two situations occurs at a Turing instability (TI) \cite{turing_alan_mathison_chemical_1952}. 
 According to this principle the bare soil state $A_0$ is always stable (unstable) against homogeneous perturbations ($k=0$) for $\alpha<0$ ($\alpha>0$).
 \begin{figure}[t]
 	\centering
 	\includegraphics[scale=1]{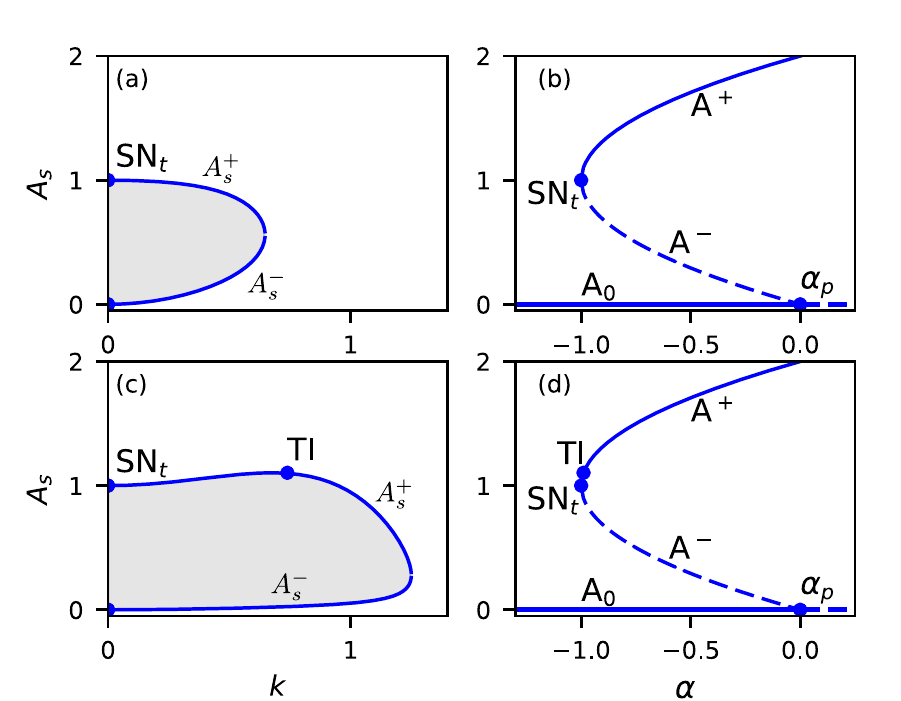}
 	\caption{(Color online) Panels (a)-(b) show the marginal instability curve and the bifurcation diagram associated with the HSS solution for $(\beta,D)=(2,1.5)$. The gray area in (a) shows the range of $A_s$ where the HSS is unstable, and corresponds to the dashed lines plotted in (b). The HSS solution is stable outside this region as shown with solid lines in (b). Panels (c)-(d) show the same type of diagrams but for $(\beta,D)=(2,0.5)$. The TI occurs at the maximum of this curve and is signaled with a blue dot in (d).}
 	\label{F_Marginal}
 \end{figure}

We analyze the stability of the uniform vegetation states $A^+$ and $A^-$ by 	
means of the marginal instability curve $A_s(k)$ [see Fig.~\ref{F_Marginal}(a),(c)]. This curve defines the band of unstable wave-numbers, i.e. modes, and is composed by the two branches $A^\pm_s(k)$,
solutions of the equation 
\begin{equation}\label{lin_sta_A}
2A_s^2+ (k^4-k^2-\beta)A_s+Dk^2=0,
\end{equation}
which is obtained by taking $\sigma=0$ in Eq.~(\ref{sigma}). 
Thus, $A_s$ is unstable to perturbations with a fixed $k$, if $A_s$ lays inside the curve, i.e. $A_s^-(k)<A_s<A_s^+(k)$, and stable otherwise.

The maximum of this curve satisfies the condition $d\sigma/dk|_{k_c}=0$, what leads to the critical wave-number
\begin{equation}\label{critical_k}
k^2_c\equiv\frac{A_c-D}{2 A_c}.
\end{equation}
Note that, in order for $k_c$ to be real, it is necessary that $D<A_c$. Then,
 the TI corresponds to the real solution $A_c$ of $\sigma|_{k_c}=0$, i.e., solution of the equation
\begin{equation}\label{hom_solution}
8A_c^3-(4\beta+1)A_c^2+2DA_c-D^2=0,
\end{equation}
which reads    
\begin{equation}\label{expAC}
A_c=\frac{1}{24}\left(1+4\beta+l_1(\beta,D)+\frac{(1+4\beta)^2-48D}{l_1(\beta,D)}\right),
\end{equation}
with
$$l_1=(1+48\beta^2+64\beta^3+\beta(12-288D)-72D+864D^2+l_2)^{1/3}$$
and 
$$l_2=48\sqrt{3}D\sqrt{\beta(1+4\beta)^2-72\beta D+2D(54D-1)}.$$

Figures~\ref{F_Marginal}(a)-(b) show the marginal stability curve and the HSS solutions for $(\beta,D)=(2,1.5)$. The maximum of this curve occurs at $k=0$, and therefore $A^+$ is stable all the way until SN$_t$ [see solid line in Fig.~\ref{F_Marginal}(b)], while $A^-$ is always unstable. Decreasing $D$, the maximum of the curve shifts from $k=0$ to a finite value $k_c$ where the TI occurs. This is the situation shown in Figs.~\ref{F_Marginal}(c)-(d) for $(\beta,D)=(2,0.5)$. Now, $A^+$ is stable for $A^+>A_c$ and unstable otherwise.  

The TI defines the manifold $\alpha_c=A_c^2-\beta A_c$ in the parameter space, and region II can be subdivided as follows:
\begin{itemize}
	\item II$_a$: $A^+$ is unstable against non-uniform perturbations ($k\neq0$). This region is defined by $\alpha_t < \alpha < \alpha_c$ when $\alpha_c$ exists.
	\item II$_b$: $A^+$ is stable against non-uniform perturbations ($k\neq0$). This region is defined by $ \alpha > \alpha_c$ when $\alpha_c$ exists, or by $ \alpha > \alpha_t$ when $\alpha_c$ does not.

\end{itemize}

  \begin{figure}[!t]
  	\centering
  	\includegraphics[scale=1]{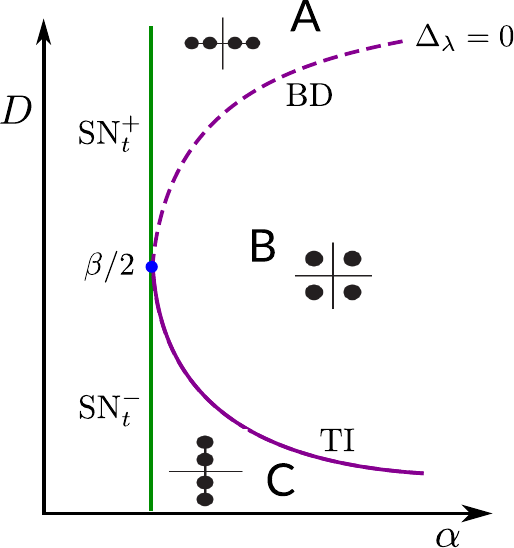}
  	\caption{(Color online) Phase diagram in the $(\alpha,D)-$parameter space showing the organization of the spatial eigenvalues $\lambda$ satisfying the equation (\ref{eq_spa_eigen_MT}) for a fixed value of $\beta$.}
  	\label{eigenvalue_conf}
  \end{figure}

\subsection{Spatial stability of the uniform states}
The steady states of the system (\ref{normal}) can be described in the framework of what is called
{\it spatial dynamics} \cite{champneys_homoclinic_1998,haragus_local_2011}. This technique consists in recasting the stationary version of Eq.~(\ref{normal}) into a finite dimensional dynamical system where the evolution in time is substituted by an evolution in space. In this context, an analogy between the solutions of the physical system (\ref{normal}), and those of the new dynamical system is established such that a uniform state of Eq.~(\ref{normal}) corresponds to a fixed or equilibrium point of the dynamical system; a LS bi-asymptotic to the HSS $A_s$ corresponds to a {\it homoclinc orbit} to a fixed point; and a front connecting two different uniform states is seen as a {\it heteroclinic orbit} between two different equilibria.

Thus, the linear dynamics near the different fixed points of the spatial system (determined by the spatial eigenvalues), and the bifurcations that they may undergo are essential to understand the origin of the different dissipative states emerging in the system \cite{champneys_homoclinic_1998,haragus_local_2011}. 

To calculate the spatial eigenvalues, it is enough to substitute $k=-i\lambda$ in Eq.~(\ref{sigma}) and find its roots, i.e. to solve:
\begin{equation}\label{eq_spa_eigen_MT}
	A_s\lambda^4+(A_s-D)\lambda^2+3A_s^2-2\beta A_s-\alpha=0.
\end{equation}
For the stable bare soil solution ($A_s=0$), Eq.~(\ref{eq_spa_eigen_MT}) becomes
\begin{equation}\label{eq_spa_eigen_MT0}
D\lambda^2+\alpha=0,
\end{equation}
which leads to the solution $\lambda_\pm=\pm\sqrt{-\alpha/D}$. For $\alpha<0$ this solution shows that the bare soil cannot have any kind of oscillations, damped or not. This is in agreement with the fact that biomass cannot be negative.

For the uniform vegetation $A^+$, the solution of Eq.~(\ref{eq_spa_eigen_MT}) becomes
\begin{equation}\label{eigen_MTAs}
\lambda=\pm\sqrt{\frac{ D-A^+ \pm \sqrt{\Delta_{\lambda}} } { 2A^+}},
\end{equation}
where 
\begin{equation}\label{eigen_Delta_MTAs}
\Delta_{\lambda}= (D-A^+)^2+4{A^+}(\alpha+2\beta A^+-3A^{+2}).
\end{equation}
Note that the condition $\Delta_{\lambda}=0$ leads to the same equation as (\ref{hom_solution}), which defines the position of the TI at $A^+=A_c$.

Depending on the control parameters of the system the spatial eigenvalues change as indicated in the $(\alpha,D)-$ phase diagram shown in Fig.~\ref{eigenvalue_conf} for a fixed value of $\beta$. The transition between these configurations occurs through the TI and the SN$_t$ bifurcations, in solid green and purple respectively, and through the dashed purple line.

The line corresponding to the fold SN$_t$ is given by Eq.~(\ref{position_SN}) and is constant for any value of $D$. Along this line the spatial eigenvalues read
\begin{equation}
	\lambda_{1,2}=0, \hspace{1cm} \lambda_{3,4}=\pm\sqrt{\frac{|2D-\beta|}{\beta}},
\end{equation}
if $2D-\beta>0$, and
\begin{equation}
\lambda_{1,2}=0, \hspace{1cm} \lambda_{3,4}=\pm i\sqrt{\frac{|2D-\beta|}{\beta}},
\end{equation}
if $2D-\beta<0$. We refer to the fold as SN$_t^+$ in the first configuration, and as SN$_t^-$ in the second. 
\begin{figure*}[t]
	\centering
	\includegraphics[scale=0.62]{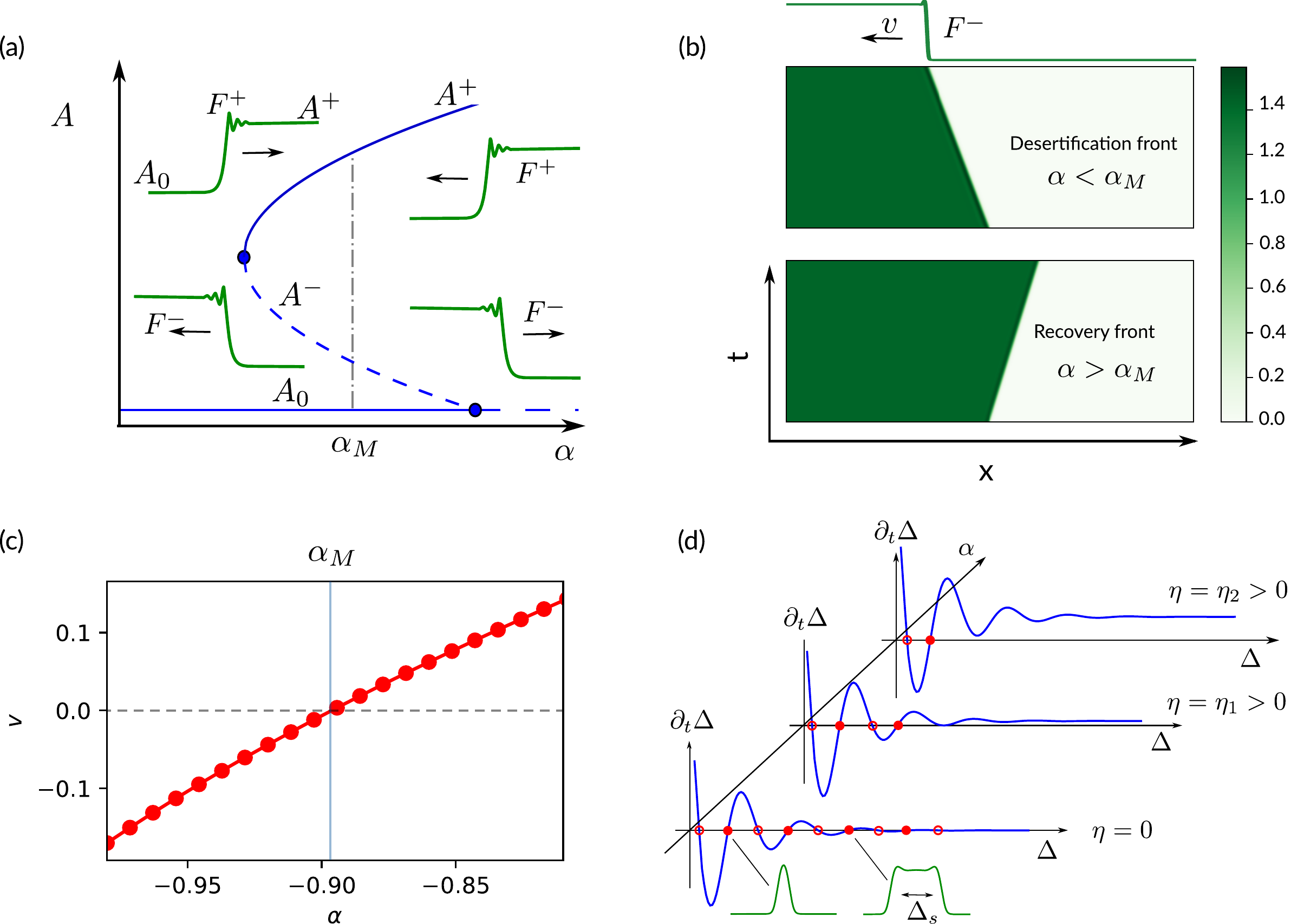}
	\caption{(Color online) (a) shows a sketch of the uniform states in a bistable configuration with the pointed-dashed gray line marking the Maxwell point of the system $\alpha_M$. Solid (dashed) lines represent stable (unstable) states. For $\alpha<\alpha_M$, $F^+$ and $F^-$ are desertification fronts; whereas for $\alpha_M<\alpha<\alpha_p$, are recovery fronts. (b) shows the temporal evolution of a desertification front for $\alpha=\alpha_M-0.08$ [top] and a recovery front for $\alpha=\alpha_M+0.08$ [bottom], when we set $(\beta,D)=(2,0.5)$. Panel (c) shows the velocity of the fronts as a function of $\alpha$ for $(\beta,D)=(2,0.5)$. The transition between positive and negative velocity occurs at the Maxwell point of the system $\alpha_M\approx-0.8962$. (d) shows a sketch of the oscillatory interaction defined by Eq.~(\ref{F_interact}) for three values of $\eta$, namely at the Maxwell point $\eta=0$, and above it ($\eta=\eta_1,\eta_2$). The stable (unstable) separations $\Delta_s$ are labeled using ${\color{red}\bullet}$ (${\color{red}\circ}$).}
	\label{Fronts_locking}
\end{figure*}
The condition $\Delta_\lambda=0$ defines the dashed and solid purple lines in Fig.~\ref{eigenvalue_conf}. The solid line corresponds to the TI with the double pure imaginary eigenvalues
\begin{equation}
\lambda_{1,2}=i\sqrt{\frac{|D-A_c|}{2A_c}}, \hspace{1cm} \lambda_{3,4}=-i\sqrt{\frac{|D-A_c|}{2A_c}},
\end{equation}
while on the dashed purple line the eigenvalues are purely real
\begin{equation}
\lambda_{1,2}=\sqrt{\frac{|D-A_c|}{2A_c}}, \hspace{1cm} \lambda_{3,4}=-\sqrt{\frac{|D-A_c|}{2A_c}}.
\end{equation}
This line corresponds to a Belyakov-Devaney transition \cite{champneys_homoclinic_1998,haragus_local_2011}, and hereafter we label it as BD.

In Fig.~\ref{eigenvalue_conf} we observe that the configuration of the
different eigenvalue is modified while crossing the previous lines.
We can classify them in three main groups:
\begin{itemize}
	\item {\bf A.} The four eigenvalues $\lambda$ are real. This region is located between SN$_t^+$ and the BD line. Trajectories approach or leave the uniform vegetation state monotonically with $x$.
	\item {\bf B.} The eigenvalues $\lambda$ consist in a quartet of complex numbers. This region is located in-between the TI and BD lines. Here the trajectories suffer a damped oscillatory dynamics in $x$ near the fixed point $A^+$.
	\item {\bf C.} The four eigenvalues $\lambda$ are imaginary. This configuration exist in-between the TI and the SN$_t^-$ lines. Here the uniform vegetation state is unstable to periodic patterns.
\end{itemize}

In Sec.~\ref{sec:3} we show that LSs can form through the locking of fronts when the dynamics near $A^+$ is described by the configuration {\bf B}, and therefore in the following we focus on this region. Furthermore, the theory of dynamical systems predicts that small amplitude LSs (i.e. homoclinic orbits) bifurcate from the TI and SN$_t^+$ \cite{champneys_homoclinic_1998,haragus_local_2011}. In Sec.~\ref{sec:4} we analytically obtain weakly non-linear LS solutions near these points. 

\section{Vegetation fronts and locking}\label{sec:3}
In this section we introduce the concept of front locking, also called pinning, as the mechanism responsible for  the formation of LSs. In region II$_b$ the system is bi-stable, i.e. the bare soil state $A_0$ and the uniform vegetation state $A^+$ coexist for the same range of parameters. 
Considering this situation, vegetation fronts may arise connecting $A_0$ with $A^+$ ($A_0\rightarrow A^+$) or vice-versa  ($A_+\rightarrow A_0$). In what follows we refer to the former front as $F^{+}$, and the later as $F^{-}$ [see Fig.~\ref{Fronts_locking}(a)]. Along this section we can have in mind infinite domains or very large finite domains.

 Either $F^+$ or $F^-$ drifts at a constant velocity $v$ that can be positive or negative depending on the control parameters of the system. These fronts are solutions of the stationary equation:
\begin{equation}
\alpha A+\beta A^2-A^3+v\partial_xA+D\partial_x^2 A-A\partial_x^2A-A\partial_x^4A=0,
\end{equation}	
that we obtain writing Eq.~(\ref{normal}) in the moving reference frame at constant velocity $v$ (i.e. considering the transformation $x\rightarrow x-vt$) and setting $\partial_tA=0$.

Furthermore, to preserve the symmetry $x\rightarrow-x$ the velocity of the fronts $F^+$ and $F^-$ must have same modulus but opposite direction. We have to notice that this dynamics is not valid, for example, in slightly sloped topographies where the previous invariance is destroyed \cite{carter_traveling_2018,bastiaansen_stable_2019}.

 The threshold between these two situations is marked with a vertical pointed-dashed gray line [see Fig.~\ref{Fronts_locking}(a)], and corresponds to the Maxwell point of the system $\alpha_M$ \cite{coullet_localized_2002}. At the Maxwell point the velocity of the front cancels out, and the front changes the direction of propagation. On the left of $\alpha_M$ the bare soil state $A_0$ invades the uniform vegetation one $A^+$, as illustrated in Fig.~\ref{Fronts_locking}(b)[top] for $F^-$ and $(\beta,D)=(2,0.5)$. Such a front is  called {\it desertification front}. In contrast, on the right of $\alpha_M$, $A^+$ invades $A_0$ [see the bottom panel in Fig.~\ref{Fronts_locking}(b)], and we refer to this front as {\it recovery front}. Hence, the Maxwell point of the system appears to be of great importance to predict desertification.  In Fig.~\ref{Fronts_locking}(c) we show the modification of the front velocity with $\alpha$ for $(\beta,D)=(2,0.5)$, which has been obtained through direct numerical simulations. For this set of parameters the Maxwell point is situated at $\alpha_M\approx-0.8962$, and is marked with a solid-blue vertical line.  
 
The tails of the front can be described asymptotically around $A^+$ by the ansatz  
\begin{equation}
A(x)-A_+\sim{\rm cos}(Kx)e^{-Qx},
\end{equation}
where $Q$ and $K$ correspond to the real and imaginary parts of the spatial eigenvalue $\lambda$. In region {\bf A} [see Fig.~\ref{eigenvalue_conf}] $K=0$, and the tails are monotonic around $A^+$, whereas in region {\bf B}, $K$ and $Q$ are different from zero, and the tails of the front show damped oscillations in $x$ around the uniform state $A^+$.

In the last case, two fronts with opposite polarity, i.e. $F^+$ and $F^-$, separated by a distance $\Delta$ experience an interaction described by
\begin{equation}\label{F_interact}
\partial_t\Delta=\varrho {\rm cos}(K \Delta)e^{-Q\Delta}+\eta,
\end{equation} 
where $\eta\sim\alpha-\alpha_M$, and therefore proportional to the separation from the Maxwell point, and $\varrho$ depends on the parameters of the system. Equation~(\ref{F_interact}) is generic and has been obtained through perturbation analysis in different systems \cite{coullet_nature_1987,coullet_localized_2002,clerc_patterns_2005,clerc_analytical_2010,escaff_localized_2015}. The interested reader can find a detailed derivation of such type of equations in Ref.~\cite{clerc_analytical_2010}.

The presence of $K\neq0$ is responsible for the oscillatory nature of the interaction which alternates attraction with repulsion, as shown in Fig.~\ref{Fronts_locking}(d) for three different values of $\eta$.  When $\partial_t\Delta=0$, the fronts lock at different stationary separations $\Delta_s$ satisfying $\varrho {\rm cos}(K\Delta_s)e^{-Q\Delta_s}+\eta=0$.
At $\alpha=\alpha_M$ ($\eta=0$) [see the bottom graph in Fig.~\ref{Fronts_locking}(d)] the width of the LSs $\Delta_s$ is quantized $\Delta_s^n=\frac{\pi}{2K}(2n+1)$, with $n=1,2,3,\dots$ \cite{coullet_nature_1987,coullet_localized_2002}. 
By increasing $n$ by one, an extra spatial oscillation or dip is nucleated in the LS. The stable (unstable) separation distances are marked with $\bullet$ ($\circ$).

As soon as $\eta>0$ the blue curve is shifted upwards (downwards if $\eta<0$), and as a result the number of stationary intersections decreases (see middle and top graphs for $\eta=\eta_1$ and $\eta_2$). Thus, the separation from $\alpha_M$ implies the disappearance of wider LSs, until eventually even the single peak LS disappears.    
In the coming sections we will see that the interaction described by Eq.~(\ref{F_interact}) is responsible for the bifurcation structure that the LSs undergo.

\section{Weakly non-linear localized states}\label{sec:4}
In Sec.~\ref{sec:3} we have introduced the mechanism of front locking to explain the formation of high amplitude LSs of different widths. 
However, this mechanism does not explain the origin of these structures from a bifurcation point of view.
Normal form theory predicts the existence of small amplitude LSs emerging from the main local bifurcations that the HSS undergoes \cite{champneys_homoclinic_1998,haragus_local_2011}.
Here, we use multi-scale perturbation theory (see Appendix~\ref{WNL}) to compute weakly nonlinear steady solutions of Eq.~(\ref{normal}) near the main bifurcations of interest: the transcritical bifurcation occurring at $\alpha_p$, the fold or SN bifurcation at $\alpha_t$, and the TI located at $\alpha_c$. 
In the neighborhood of such bifurcations, weakly nonlinear states are 
captured by the ansatz:

$$	A(x,t)-A_s\sim \epsilon a(X)e^{ik_cx}+c.c.,$$

where $\epsilon\ll1$ measures the onset from the bifurcation, $k_c$ is the characteristic wave-number of the marginal mode at the bifurcation ($k_c=0$ for the fold and transcritical, and $k_c\neq0$ for the TI) and $a$ is the amplitude or envelope describing a modulation occurring at a larger scale $X=\epsilon^lx$, with the election of $l$ depending on the problem. 
In what follows we show the analytical expressions for small amplitude states about the different bifurcations of the system, and refer to Appendix~\ref{WNL} for a detailed exposition of the analysis. 
We have to point out that the temporal stability of such asymptotic states can be estimated analytically as done in \cite{kolokolnikov_existence_2005,van_heijster_pulse_2008,chen_stability_2011}. However, here, the temporal stability is calculated numerically (see Sec.~\ref{sec:5}).

\subsection{Small amplitude spots around the transcritical bifurcation}
In Appendix~\ref{WNL_Trans} we show that near $\alpha_p$ small amplitude spots of the form
\begin{equation}\label{small_spot}
A(x)=-\frac{3}{2}\frac{\alpha}{\beta}{\rm sech}^2\left(\frac{1}{2}\sqrt{\frac{-\alpha}{D}}x\right),
\end{equation}
exist for negative values of $\alpha$.
Figure~\ref{agreement_LSs}(a) shows the profile of such structure in blue. The red dashed line corresponds to the exact numerical solution, that has been obtained using a Newton-Raphson solver and considering (\ref{small_spot}) as initial guess. The plot shows excellent agreement.
\subsection{Small amplitude gaps around the fold bifurcation}
\begin{figure}[t]
	\centering
	\includegraphics[scale=1]{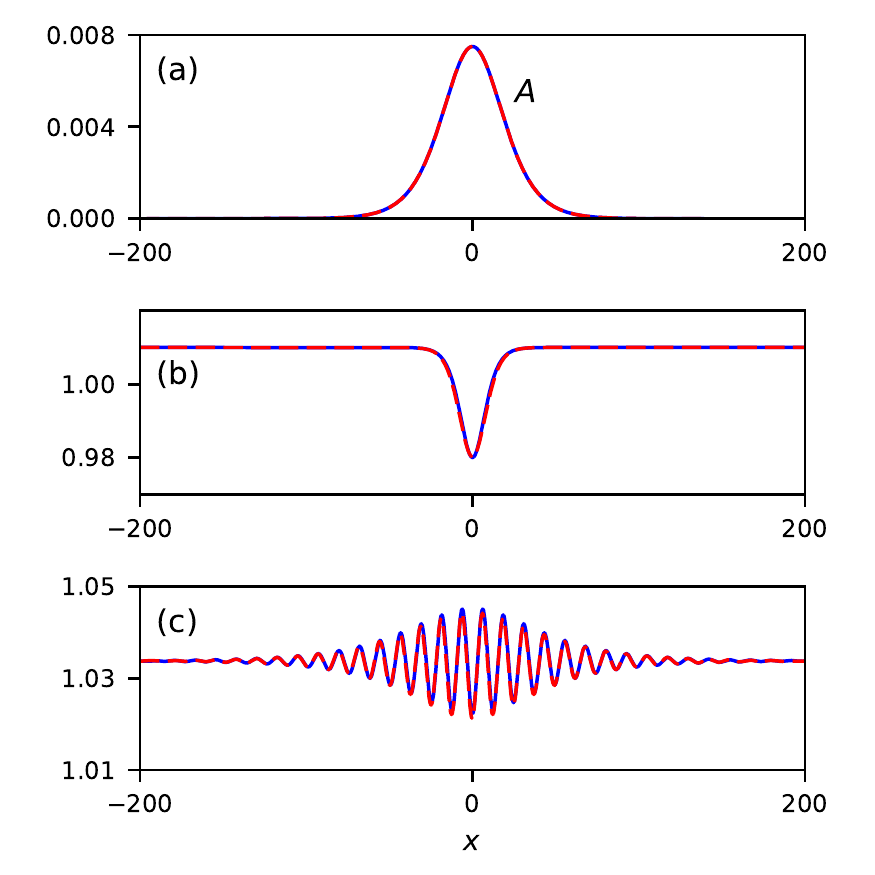}
	\caption{(Color online) (a) shows the weakly nonlinear spot solution (\ref{small_spot}) using a solid blue line, and the exact numerical solution (dashed red line) obtained from a Newton-Raphson solver for $D=1.5$ and $\alpha=\alpha_p-0.01$; (b) shows in blue the analytical gap solution (\ref{small_gap1}) and the exact numerical one (dashed red) for $D=1.5$ and $\alpha=\alpha_t+0.0001$; (c) shows the agreement for the bump solution arising from the TI for $D=0.5$ and $\alpha=\alpha_c+0.00003$. For every case we choose $\beta=2$.}
	\label{agreement_LSs}
\end{figure}

Following a similar procedure, in Appendix~\ref{WNL_Fold} we show that in a neighborhood of $\alpha_t$ small amplitude gap states of the form 
\begin{multline}\label{small_gap1}
A(x)=\frac{\beta}{2}+\sqrt{\alpha-\alpha_t}\\-3\sqrt{\alpha-\alpha_t}{\rm sech}^2\left(\frac{1}{2}\sqrt{\frac{2\beta}{2D-\beta}}(\alpha-\alpha_t)^{1/4}x\right),
\end{multline}
arise whenever $2D-\beta>0$. This last condition ensures that $A^+$ is stable all the way until SN$_t^+$ at $\alpha_t$.
Figure~\ref{agreement_LSs}(b) shows in solid-blue the small amplitude gap and in dashed-red the exact numerical solution obtained also from a Newton-Raphson algorithm. Like in the previous case, the plot shows very good agreement.
\subsection{Small amplitude gaps around the Turing bifurcation}
In Appendix~\ref{WNL_Turing} we perform the weakly non-linear analysis about the Turing bifurcation at $\alpha_c$.  In this case, small amplitude stationary gap periodic patterns of the form
\begin{equation}\label{pattern_gap}
A(x)=A_c+(\alpha-\alpha_c)\tilde{A}_2+2\sqrt{\frac{\alpha-\alpha_c}{-c_3}}{\rm cos}(k_c x+\varphi),
\end{equation}
arise from the TI, where $A_c$, $\tilde{A}_2$, and $c_3$ depend on the parameters of the system [see Appendix~\ref{WNL_Turing}], $k_c$ is the wave-number associated with the critical pattern emerging from the TI [see Eq.~(\ref{critical_k})], and $\varphi$ is an arbitrary phase. For the range of parameters explored in this work $c_3$ is always negative, and therefore, the periodic pattern arises subcritically. 
	
In this situation small amplitude gap states of the form 
\begin{multline}\label{LS_Turing}
A(x)=A_c+(\alpha-\alpha_c)\tilde{A}_2\\-2\sqrt{\frac{2(\alpha-\alpha_c)}{-c_3}}{\rm sech}\left(\sqrt{\frac{\alpha-\alpha_c}{-c_2}}x\right){\rm cos}(k_cx+\varphi),
\end{multline}
emerge, together with the subcritical pattern, from the TI if $c_2<0$, what occurs whenever  $2D-\beta<0$. Figure~\ref{WN} in  Appendix~\ref{WNL_Turing} shows the dependence of $c_2$ and $c_3$ with $D$ for $\beta=2$. 
\begin{figure*}[t]
	\centering
	\includegraphics[scale=1]{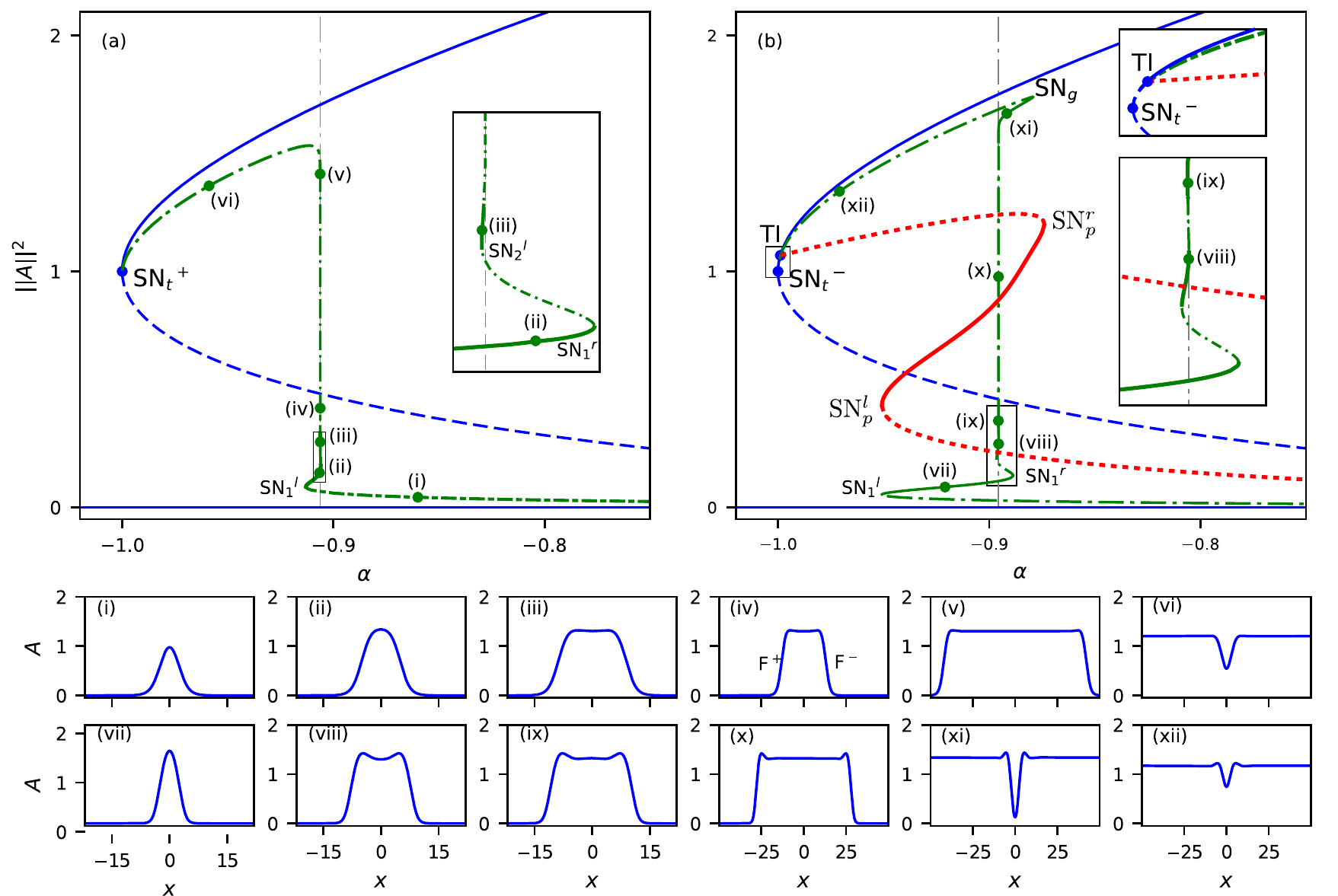}
	\caption{(Color online) Bifurcation diagrams for $\beta=2$ and two different values of $D$: $D=1.5$ in (a), and $D=0.5$ in (b). Stable (unstable) solutions branches are marked with solid (pointed-dashed lines). The green dots correspond to the solutions shown in sub-panels (i)-(xii). The red lines correspond to the gap pattern branches arising from the TI. The subpanel in (a) shows a close-up view of the collapsed snaking around $\alpha_M$. The bottom subpanel in (b) shows a similar view of the diagram, and the top panel shows the subcritical birth of the gap pattern and localized gap from the TI at $\alpha_c$.  Here we consider a domain width $L=100$ and periodic boundary conditions.}
	\label{diagram}
\end{figure*}


The phase $\varphi$ remains arbitrary within the asymptotic theory. However,  expansions beyond all orders show that two specific values of this phase are selected, namely $\varphi=0$, and $\varphi=\pi$ \cite{kozyreff_asymptotics_2006}. 

Panel \ref{agreement_LSs}(c) shows in blue the analytical solution (\ref{LS_Turing}) for $\varphi=0$, and in dashed red the exact numerical solution. The overlap shows very good agreement between both calculations. In what follows we focus on the gap states associated with $\varphi=0$.



\section{Bifurcation structure of localized states: collapsed snaking}\label{sec:5}
 
In this section we study the bifurcation structure of the different dissipative LSs arising in this system. To do so we apply numerical continuation algorithms based on a Newton-Raphson solver which allows us to track the steady states of the system in their different control parameters \cite{allgower_numerical_1990}. In this way, we are able to study how the different LSs appearing in the system are organized in terms of bifurcation diagrams. For these calculations we consider a finite domain of length $L=100$ and periodic boundary conditions.

\begin{figure}[!t]
	\centering
	\includegraphics[scale=1]{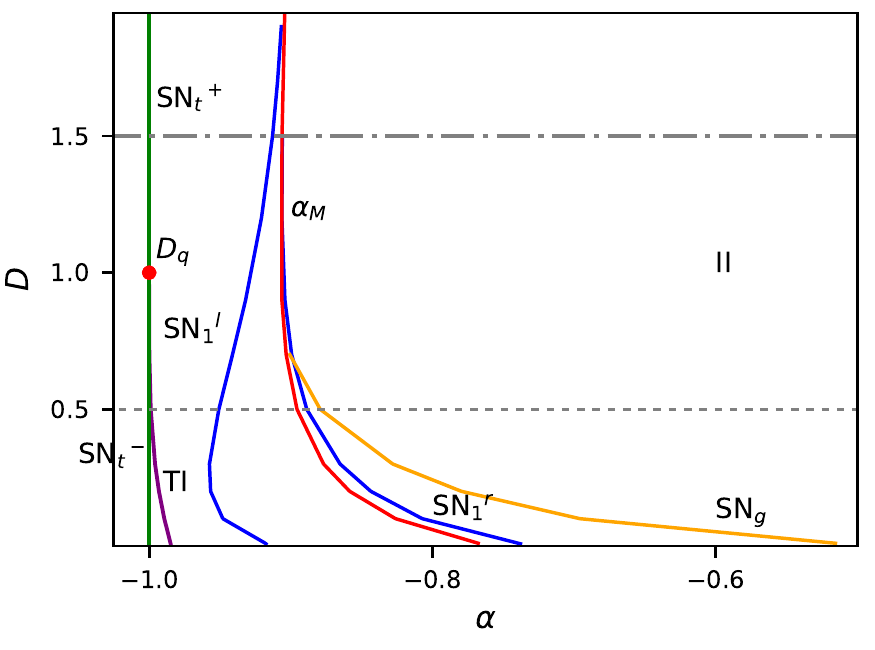}
	\caption{(Color online) Phase diagram in the $(\alpha,D)-$parameter space for $\beta=2$. The different lines correspond to the bifurcations shown in Fig.~\ref{diagram}: the fold SN$_t$ (green line), the TI (purple), the saddle-nodes SN$_1^{l,r}$ of the spot LSs (blue), the Maxwell point $\alpha_M$ (red), and the fold of the gap LSs SN$_g$ in orange. The point-dashed gray line corresponds to the bifurcation diagram show in Fig.~\ref{diagram}(a); the pointed gray line to the one shown in Fig.~\ref{diagram}(b).}
	\label{phase_1}
\end{figure}

In Sec.~\ref{sec:4} we have calculated analytically weakly nonlinear solutions corresponding to different types of pulses (spots or gaps) which are only valid in the neighborhood of the local bifurcations of the uniform state. Starting from these analytical solutions, we have calculated the bifurcation diagrams shown in Fig.~\ref{diagram} for $\beta=2$ and two representative values of $D$, namely $D=1.5$ in (a) and $D=0.5$ in (b).
These diagrams show the $L^2$-norm $$||A||^2=\frac{1}{L}\int_{-L/2}^{L/2}A(x)^2dx$$ as a function of the parameter $\alpha$. In particular, the diagrams correspond to slices for constant $D$ of the phase diagram in the $(\alpha,D)-$parameter space shown in Fig.~\ref{phase_1}, where the main bifurcation lines of the system are plotted.

For $D=1.5$ (see point-dashed gray line in Fig.~\ref{phase_1}) the situation is like the one depicted in Fig.~\ref{diagram}(a). In a neighborhood of the transcritical bifurcation $\alpha_p$ a spot solution of the form (\ref{small_spot}) exists. This solution is temporally unstable all the way until SN$_1^l$, and increases its amplitude while decreasing $\alpha$ [see profile (i)].

When an analytical expression for a LS is known, its temporal stability can be computed analytically as has been done by different authors 
\cite{chen_oscillatory_2009,kolokolnikov_existence_2005,chen_stability_2011,van_heijster_pulse_2008,doelman_nonlinear_2007,makrides_existence_2019}.
Here, however, we only have access to the LSs solutions numerically, and therefore, their stability can only be determined by solving numerically the eigenvalue problem 
\begin{equation}
	\mathcal{L}\psi=\sigma\psi,
\end{equation}
where $\mathcal{L}$ is the linear operator (\ref{lin_op}) evaluated at a given LS field $A$, and $\sigma$ and $\psi$ are the eigenvalues and eigenvectors associated with $\mathcal{L}$ at a given set of parameters.

Once the fold SN$_1^l$ is passed the spot state becomes stable [see panel (ii)] and conserves the stability until SN$_1^r$, where it becomes unstable once more.
Proceeding up in the diagram (i.e increasing $||A||^2$) the LS
broadens [see profiles (iii)-(iv)] as a result of the nucleation of new dips (i.e. spatial oscillations) at every SN$_i^r$ [see close-up view in Fig.~\ref{diagram}(a)]. In the meanwhile, the solution branches suffer a sequence of exponentially decaying oscillations around the Maxwell point $\alpha_M$, and the different solutions gain and loss stability through a series of saddle-node bifurcations SN$_i^{l,r}$. 
This type of bifurcation structure is known as {\it collapsed snaking}, and has been studied in detail in different systems \cite{knobloch_homoclinic_2005,burke_classification_2008,ma_defect-mediated_2010,parra-rivas_dark_2016}. At this stage [see profile (iv)] we can see that the LS is formed by the locking of two fronts $F^+$ and $F^-$, as described by Eq.~(\ref{F_interact}). 


In Sec.~\ref{sec:4}, we have also obtained the analytical expression (\ref{small_gap1}) for a gap solution in the neighborhood of SN$_t^+$. Tracking numerically this state to larger values of $\alpha$, the LS deepens [see profile (vi)] until reaching its maximal depth at $\alpha_M$. Along this branch, the gap is always temporally unstable. In finite domains the spot and gap types of solutions are inter-connected as shown in the bifurcation diagram of Fig.~\ref{diagram}(a). Furthermore, we have to mention that in the presence of periodic boundary conditions, one cannot properly discern between both types of states. Indeed, a spot LS [e.g. state (v)] can be seen as gap by just applying a translation of $L/2$. However, in real two-dimensional landscapes periodic boundary conditions are senseless, and spots and gaps are two different, and well defined, structures.


Figure~\ref{phase_1} shows how the main bifurcation lines of the system vary with the diffusion $D$. These lines correspond to the Maxwell point $\alpha_M$ (solid red line), the folds SN$_{t}^{\pm}$ (solid green line), the folds of the spot structures SN$_1^{l,r}$, the fold of the gap solutions SN$_g$, and the TI $\alpha_c$ (purple solid line). Decreasing $D$ the TI emerges from the fold SN$_{t}^-$ at $D=D_{q}\equiv\beta/2=1$. Bellow this point (i.e. $D<D_{q}$) $A^+$ is unstable between TI and SN$_{t}^-$, and this unstable region increases whereas decreasing $D$. For $D>D_{q}$ the TI fades away, and the uniform vegetation state $A^+$ is stable everywhere, as shown in Fig.~\ref{diagram}(a).  

Decreasing $D$ bellow $D_{q}$ the region of existence of the spot LSs becomes wider, and every branch of solutions widens. This situation corresponds to the diagram depicted in Fig.~\ref{diagram}(b) for $D=0.5$. While in Fig.~\ref{diagram}(a) the solution branches of the LSs collapse rapidly to $\alpha_M$ as increasing $||A||^2$,
in panel (b) the collapse is much slower, and therefore, branches of wider structures persist. Examples of these type of LSs are shown in panels (vii)-(x). 

The perturbative analysis of Sec.~\ref{sec:4} shows that localized small amplitude gap solutions of the form (\ref{LS_Turing}) arise from the TI , together with a subcritical periodic gap pattern [see Eq.~(\ref{pattern_gap})], whenever $c_2$ and $c_3$ are both negative. This branching behavior around the TI is shown in detail in the top close-up view of Fig.~\ref{diagram}(b), where the 
solution branches of the gap pattern are shown in red while the localized gaps in green.

The localized gap states are then unstable between $\alpha_c$ and SN$_g$, and increase their amplitude as approaching SN$_g$. An example of this gap state is plotted in panel (xii). 
Once SN$_g$ is crossed, the gap solution becomes stable [see panel (xi)], continuing stable until reaching $\alpha_M$.
In contrast to the localized gap plotted in panel (vi), the gap shown here possesses oscillatory tails around $A^+$. As proceeding down in the diagram the localized gap (xi) broadens and, in a periodic domain, eventually becomes the spot state shown in (x).

The subcritical (unstable) gap pattern emerging from the TI increases its amplitude as approaching SN$_p^r$, where it becomes stable. The stability is then preserved until reaching SN$_p^l$ where the branch of patterns folds back and eventually connects with $A^-$ [see red branches in Fig.~\ref{diagram}(b)]. 
Note that the subcritical gap pattern may be responsible for the appearance of a homoclinic snaking scenario as already reported in other works on this model \cite{zelnik_desertification_2017}. Furthermore, the transition between the collapsed and the homoclinic snaking structures may be related to the presence of isolas of localized patterns as described in \cite{zelnik_implications_2018}. However, the confirmation of this scenario requires further investigation.

The collapsed snaking structure is a consequence of the damped oscillatory interaction experienced by the two fronts [see Eq.~(\ref{F_interact})], and can be understood from the sketch shown in Fig.~\ref{Fronts_locking}(d). At the Maxwell point ($\eta=0$) a number stable and unstable LSs form at the stationary front separations $\Delta^n_s$. Stable (unstable) LSs in Fig.~\ref{Fronts_locking}(d) then correspond to a set of points on top of the stable (unstable) branches of solutions at $\alpha_M$ in the collapsed snaking diagrams of Fig.~\ref{diagram}. As the parameter $\alpha$ separates from $\alpha_M$, the branches of wider LSs, both stable and unstable, start to disappear in a sequence of fold bifurcations, and only narrow LSs survive. This scenario corresponds to the situation shown in Fig.~\ref{Fronts_locking}(d) for $\eta=\eta_1$, where the number of intersections of $\partial_t\Delta$ with zero decreases, and with it, the number of LSs. In this framework, the fold bifurcations of the collapsed snaking diagram take place when the extrema of $\partial_t\Delta$ become tangent to zero. Increasing $\alpha$ even further, less and less intersections occur [see Fig.~\ref{Fronts_locking}(d) for $\eta=\eta_2$] until eventually the last fold SN$_1^r$ is passed and the single spot destroyed.

The type of LSs studied here can be extremely useful for predicting the onset of a desertification process.  Desertification is related to the presence of the so called desertification fronts occurring for $\alpha<\alpha_M$. Hence, determining how far is the ecosystem from the Maxwell point is quite relevant.
The phase diagram shown in Fig.~\ref{phase_1} indicates that the region of existence of spots and gaps is quite localized around $\alpha_M$.  As an example, the presence of the kind of gaps studied here indicates that the ecosystem is in the recovery regime. In other words, the presence of single gaps may indicate that a full vegetated state is stable and a flat front is always increasing the biomass of the ecosystem. On the other hand, the presence of spots which come from a collapsed snaking is not strictly related with one side of the Maxwell point, but it usually indicates that the ecosystem is in the desertification region.  Indeed, considering $D=0.5$, approximately the $90\%$ of the spots are found in the desertification region, while the $10\%$ belongs to the recovery one. 

\begin{figure}[!t]
	\centering
	\includegraphics[scale=1]{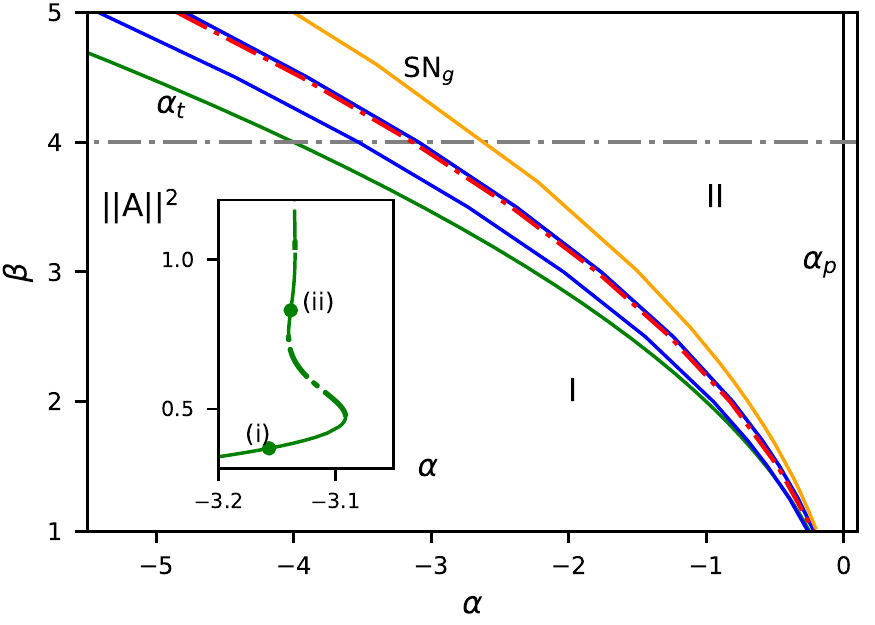}
	\includegraphics[scale=1]{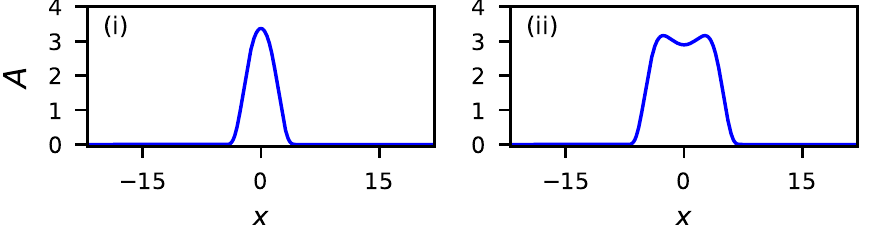}
	\caption{(Color online) Phase diagram in the $(\alpha,\beta)-$parameter space for $D=0.1$. The different lines correspond to the main bifurcations of the system: the fold SN$_t$ (green line), the TI (purple), the saddle-nodes SN$_1^{l,r}$ of the spot LSs (blue), the Maxwell point $\alpha_M$ (red), the fold of the gap LSs SN$_g$ in orange, and the transcritical bifurcation $\alpha_p$ in black.
	The point-dashed gray line corresponds to the collapsed snaking shown in the inset panel, and the labels (i)-(ii) to the LSs shown in panels bellow.}
	\label{phase_2}
\end{figure}

\section{Localized structures in the $(\alpha,\beta)-$parameter space}\label{sec:6}
In previous sections we have fixed the root-to-shoot ratio to $\beta=2$ and studied how the different types of localized gaps and spots of  vegetation, and their bifurcation structure, are modified when changing the diffusion $D$. In this section we explore how the previous scenario changes when the root-to-shoot ratio $\beta$ is varied with fixed $D=0.1$. Figure~\ref{phase_2} shows the phase diagram in the $(\alpha,\beta)-$parameter space, where the main bifurcation lines of the system are plotted.

Region II, i.e. the region of bi-stability between the bare soil and the uniform vegetation state, shrinks by decreasing $\beta$, and with it, the region of existence of spots, limited by the solid-blue lines (SN$_1^l$ and SN$_1^r$) also shrinks. Furthermore, as proceeding down in $\beta$ the different folds of the spot SN$_i^{l,r}$ approach each other and disappear in a cascade of cusp bifurcations at the Maxwell point (not shown here).

In contrast, by increasing $\beta$ the uniform vegetation state becomes more and more subcritical and the region of existence of the spot states widens. This result is closely related with the fact that 
larger root-to-shoot ratios allow plants to uptake more water from the soil, which in turn can improve their adaptation, and enlarge their stability region, as it was already shown in the context of patterns \cite{meron_nonlinear_2015}.


In most of this region, the spots undergo collapsed snaking. An example of such snaking is shown in the inset of Fig.~\ref{phase_2} for the fixed value $\beta=4$ (see pointed-dashed gray line). 
Labels (i)-(ii) correspond to the spot profiles plotted in the sub-panels bellow the phase diagram. 
For very low values of $\beta$ the region of bi-stability shrinks, and consequently the region of fronts becomes small and not relevant under small variations in rainfall.

The region between $\alpha_M$ (solid-red line) and the fold SN$_g$ (solid-orange line), where gap LSs exist, undergoes a similar behavior: it widens with increasing subcriticality, and shrinks otherwise.
  
In essence, these results show that the LSs studied in this work are robust and persist in a wide range of the parameters of the ecosystem model. 
\section{Discussion and Conclusions}\label{sec:7}
In this work we have presented a detailed study of the formation and bifurcation structure of localized vegetation spots and gaps arising in semi-arid regions close to the desertification onset, and therefore close to the Maxwell point of the system. To perform this analysis we have focus on the reduced model (\ref{normal}) in one spatial dimension, that has been derived from different models in plant ecology \cite{lejeune_model_1999,lejeune_localized_2002,tlidi_vegetation_2008,fernandez-oto_front_2019}. However, the results presented here can be relevant in the context of different pattern-forming living systems having nonviable states that undergo subcritical instabilities to viable states.

Applying multi-scale perturbation theory we have found that small amplitude spots arise from the transcritical bifurcation at $\alpha_p\equiv0$. This state increases its amplitude and eventually undergoes collapsed snaking: the spot solution branches experience a sequence of exponentially decaying oscillations around the Maxwell point of the system. As a result the localized spots, now formed by the locking of two vegetation fronts of different polarity, increase their width. Indeed the collapsed snaking is a direct consequence of the interaction of fronts as described by Eq.~(\ref{F_interact}). Due to this bifurcation structure it is much easier to find spots with a single peak than wider structures, which accumulate at parameter values very close to the Maxwell point.

Localized vegetation gaps emerge from the uniform vegetation state at two different points depending on the value of $D$. For $D>D_{q}$ they arise unstably from the uniform vegetation fold SN$_t$, and in a periodic domain, connect back with the spot states at $\alpha_M$. In contrast, for $D<D_{q}$ the situation is rather different. In this case localized gaps arise together with a periodic vegetation pattern from a Turing instability. The gaps arise initially unstable but stabilize in the fold SN$_g$. As before, these states connect back with the spots at $\alpha_M$. In principle these gap states could undergo homoclinic snaking; however, for the regime of parameters explored here, such structure has not been found.

We have also classified the different type of LSs, both spots and gaps, in two phase diagrams in the $(\alpha,D)-$ and $(\alpha,\beta)-$parameter space. For a constant $\beta$, the  $(\alpha,D)-$phase diagram shows how, the region of existence of both types of LSs widens as $D$ is decreased, and shrinks otherwise. Fixing $D$, the $(\alpha,\beta)-$phase diagram now shows how the region of existence enlarges with increasing $\beta$, and therefore with the subcriticality of the uniform vegetation state. The enlargement of the LSs stability region  may be related with the improvement of plants capacity for uptaking water for large root-to-shoot ratios, as it is the case in the context of vegetation patterns \cite{meron_nonlinear_2015}.
 
The LSs presented here are robust, and persist for a large range of parameters. These states arise in the proximity of the Maxwell point of the system, which signal the threshold between desertification and recovering processes. The presence of gaps in a ecosystem strongly suggest that the system is in a recovery region, as they only exist $\alpha>\alpha_M$.  Therefore, any flat front expands and covers the bare soil with vegetation. In contrast, spots exist, in approximately a 90$\%$, for $\alpha<\alpha_M$, and therefore their presence may indicate that the system is in the desertification region. Thus, any flat front connecting the bare soil with homogeneous vegetation may end up, with large probability, in a completely unproductive ecosystem.

We plan to study the dynamics and bifurcation structure of spots and gaps in two spatial dimensions. In this context the interaction of vegetation fronts is much more complex due to the effect of the front curvature and the presence of front instabilities which are not present in one spatial dimension \cite{hagberg_complex_1994,gomila_stable_2001,hagberg_linear_2006,fernandez-oto_front_2019}. 
	
Another interesting line of research is to understand how spots and gaps evolve under perturbations of the ecosystem such as periods of weak droughts. In this context, close to $\alpha_M$  one could expect that isolated gaps which are  temporally perturbed slightly bellow $\alpha_M$ (e.g. a weak drought) may trigger gradual desertification, as the system is brought momentarily to the desertification region. 
In contrast as the region of stability of spots is larger, weak droughts usually do not generate gradual desertification, whereas  strong droughts may imply abrupt desertification.

	
The ultimate hope is that studies of this kind will prove useful for understanding the dynamics and stability of LSs in pattern forming systems, in particular in the context of plant ecology.

 \acknowledgments 
 
 We acknowledge the financial support of Fonds de la Recherche Scientifique F.R.S.-FNRS, (P.P.-R.) and FONDECYT Project No. 3170227. (C.F.-O.). The authors are grateful to M. Chia, P. Gandhi and J. Cisternas for their useful comments during the writing of the manuscript. 


\appendix


\section{WEAKLY NON-LINEAR ANALYSIS}\label{WNL}
In this appendix, we calculate stationary weakly nonlinear dissipative
structures using multiple scale perturbation theory near the main bifurcations of the HSS of the system, namely the transcritical, fold and Turing bifurcations.

To study these types of solutions we consider the ansatz
\begin{equation}
A(x,t)=A_s+u(x),	
\end{equation}
to decouple Eq.~(\ref{normal}) in the equation for the homogeneous state $A_s$ 
\begin{equation}\label{decoupled_hom}
-A_s^3+\beta A_s^2+\alpha A_s=0,
\end{equation}
and the stationary equation for the space-dependent component $u(x)$, namely
\begin{equation}\label{decoupled_space}
(\mathcal{L}+\mathcal{N})u=0,
\end{equation}
where $\mathcal{L}$ and $\mathcal{N}$ are the linear and non-linear operators 
\begin{subequations}
	\begin{equation}\label{lin_opb}
	\mathcal{L}\equiv \alpha+ 2\beta A_s-3A_s^2+D\partial_x^2-A_s\partial_x^2-A_s\partial_x^4
	\end{equation}
	\begin{equation}
	\mathcal{N}\equiv -u^2+(\beta-3 A_s)u-u\partial_x^2-u\partial_x^4.
	\end{equation}
\end{subequations}
Following \cite{burke_classification_2008,parra-rivas_dark_2016,parra-rivas_bifurcation_2018}, we fix the value of both $D$ and $\beta$, consider $\alpha$ as the bifurcation parameter, and for each case, we introduce appropriate asymptotic expansions for $A_s$, $u$, and $x$ in terms of $\epsilon$.
In what follows we show the detail calculation for each of the cases considered in this manuscript.
\subsection{WEAKLY NON-LINEAR STATES NEAR THE TRANS-CRITICAL BIFURCATION}\label{WNL_Trans}

The transcritical bifurcation occurs at $A_s=A_0\equiv0$ at the parameter value $\alpha_p$, and the solution of the system reduces to $A(x,t)=u(x,t)$. In this particular case 
the linear and nonlinear operators read 
\begin{subequations}
	\begin{equation}
	\mathcal{L}\equiv \alpha+D\partial_x^2,
	\end{equation}
	\begin{equation}
	\mathcal{N}\equiv -u^2+\beta u-u\partial_x^2-u\partial_x^4.
	\end{equation}
\end{subequations}
In this case an appropriate asymptotic expansion for the control parameter $\alpha$ as a function of the expansion parameter $\epsilon$ is $\alpha=\alpha_p+\epsilon\delta$, whereas the space dependent variable $u$ can be expanded as
\begin{equation}
u(X)=\epsilon u_1(X)+\epsilon^2 u_2(X)+\epsilon^3 u_3(X)+\cdots,
\end{equation}
where any variable depends on the long scale variable $X\equiv\sqrt{\epsilon}x$. In this way the differential operator becomes $\partial_x^2=\epsilon\partial^2_X$.

With these considerations the linear operator expands as $\mathcal{L}=\mathcal{L}_0+\epsilon\mathcal{L}_1+\mathcal{O}(\epsilon^3)$, with  
\begin{equation}
\mathcal{L}_0=\alpha_p,\hspace{0.5cm} \mathcal{L}_1=\delta+D\partial_X^2,
\end{equation}
while the nonlinear operator develops as $\mathcal{N}=\epsilon\mathcal{N}_1+\mathcal{O}(\epsilon^2)$,
with 
\begin{equation}
\mathcal{N}_1=\beta u_1. 
\end{equation}
Inserting the previous expansions in Eq.~(\ref{decoupled_space}), and matching the different terms at the same order in $\epsilon$ we get the next two equations at order $\epsilon$ and $\epsilon^2$:
\begin{subequations}
	\begin{equation}
	\mathcal{O}(\epsilon):\hspace{0.5cm}\mathcal{L}_0 u_1=0,
	\end{equation}
	\begin{equation}\label{expans_trans_2}
	\mathcal{O}(\epsilon^2):\hspace{0.5cm}\mathcal{L}_0 u_2+\left(\mathcal{L}_1+\mathcal{N}_1\right)u_1=0.
	\end{equation}
\end{subequations}
The solvability condition at $\mathcal{O}(\epsilon)$ implies that $\alpha_p=0$ for a solution $u_1=a(X)$.

At $\mathcal{O}(\epsilon^2)$ Eq.~(\ref{expans_trans_2}) reduces to 
\begin{equation}
\left(\mathcal{L}_1+\mathcal{N}_1\right)u_1=0,
\end{equation}
what leads to the amplitude equation

\begin{equation}\label{Ampli_trans}
c_1\partial_X^2 a(X)+c_2 a(X)+c_3 a(X)^2=0,
\end{equation}
with
\begin{equation}
c_1=D,\hspace{0.4cm}	c_2=\delta,\hspace{0.4cm}c_3=\beta.
\end{equation}
This amplitude equation has two homogeneous solutions
\begin{equation}
a(\delta+\beta a)=0,
\end{equation}
as corresponds to a transcritical bifurcation, and furthermore supports
pulse solutions of the form
\begin{equation}
a(X)=-\frac{3}{2}\frac{c_2}{c_3}{\rm sech}^2\left(\frac{1}{2}\sqrt{\frac{-c_2}{c_1}}X\right),
\end{equation}
i.e. 
\begin{equation}
a(X)=-\frac{3}{2}\frac{\delta}{\beta}{\rm sech}^2\left(\frac{1}{2}\sqrt{\frac{-\delta}{D}}X\right).
\end{equation}

Thus, we conclude that in a neighborhood of $\alpha_p$ small amplitude spots of the form
\begin{equation}\label{small_spotb}
A(x)=-\frac{3}{2}\frac{\alpha}{\beta}{\rm sech}^2\left(\frac{1}{2}\sqrt{\frac{-\alpha}{D}}x\right)+\mathcal{O}(\alpha^2),
\end{equation}
exist for negative values of $\alpha$.

\subsection{WEAKLY NON-LINEAR STATES NEAR THE FOLD BIFURCATION}\label{WNL_Fold}
	Here we perform weakly non-linear analysis about the fold bifurcation of the uniform state SN$_t$ occurring at $A_t$. In this case a proper asymptotic expansion for the control parameter $\alpha$, and the uniform and space-dependent variables $A_s$ and $u$, about the fold reads:
	\begin{subequations}
		\begin{equation}
		\alpha=\alpha_t+\delta \epsilon^2,	
		\end{equation}
		\begin{equation}\label{asymp_hom_fold}
		A_s=A_t+\epsilon A_1+\epsilon^2 A_2+\cdots,
		\end{equation}
		\begin{equation}\label{asymp_space_fold}
		u(X)=\epsilon u_1(X)+\epsilon^2 u_2(X)+\cdots,	
		\end{equation}
	\end{subequations}
where the space dependent variables are functions of the long scale $X=\sqrt{\epsilon}x$. In what follows we first solve the homogeneous problem (\ref{decoupled_hom}), and later Eq.~(\ref{decoupled_space}).
\subsubsection*{\bf Solution of the uniform problem}
Inserting the asymptotic expansion (\ref{asymp_hom_fold}) 
in Eq.~(\ref{decoupled_hom}) we derive the following set of linear equations for the uniform state:
\begin{subequations}
	\begin{equation}
	\mathcal{O}(\epsilon^0):\hspace{0.5cm} (-A_t^2+\beta A_t+\alpha_t) A_t=0,	
	\end{equation}
	\begin{equation}\label{def_L0}
	\mathcal{O}(\epsilon^1):\hspace{0.5cm} \mathcal{L}_0 A_1\equiv(-3 A_t^2+2\beta A_t+\alpha_t)A_1=0,	
	\end{equation}
	\begin{equation}
	\mathcal{O}(\epsilon^2):\hspace{0.5cm} \mathcal{L}_0A_2-3A_tA_1^2+\beta A_1^2+\delta A_t=0.
	\end{equation}
\end{subequations}
The solutions at $\mathcal{O}(\epsilon^0)$ gives
\begin{equation}\label{alpha_t_aa}
\alpha_t=A_t^2-\beta A_t.
\end{equation}
The equation at $\mathcal{O}(\epsilon^1)$, has a non-trivial solution (i.e. $A_1\neq0$) if $\mathcal{L}_0=0$, from where one obtains
\begin{equation}\label{alpha_t_b}
\alpha_t=3 A_t^2-2\beta A_t.
\end{equation}
From conditions (\ref{alpha_t_aa}) and (\ref{alpha_t_b}) one finally gets the value of $A$ at the fold,
\begin{equation}
A_t=\frac{\beta}{2}.
\end{equation}
The solution $A_1$ is obtained from the solvability condition at $\mathcal{O}(\epsilon^2)$, namely 
\begin{equation}
-3A_tA_1^2+\beta A_1^2+\delta A_t=0,
\end{equation}
from where one obtains
\begin{equation}
A_1=\pm\sqrt{\frac{\delta A_t}{3A_t-\beta}}=\pm\sqrt{\delta}.
\end{equation}

\subsubsection*{\bf Solution of the space dependent problem}
Inserting the asymptotic expansion (\ref{asymp_space_fold}) 
the linear operator (\ref{lin_opb}) becomes $\mathcal{L}=\mathcal{L}_0+\epsilon\mathcal{L}_1+\mathcal{O}(\epsilon^2),$ with $\mathcal{L}_0$ defined previously in Eq.~(\ref{def_L0}), and 
	\begin{equation}
	\mathcal{L}_1=2\beta A_1-6A_t A_1+(D-A_t)\partial_X^2,
	\end{equation}
and the expansion for the nonlinear operator 
$\mathcal{N}=\epsilon\mathcal{N}_1+\mathcal{O}(\epsilon^2)$,
with 
\begin{equation}
\mathcal{N}_1=(\beta-3A_t)u_1.
\end{equation}
The insertion of the previous expansions in Eq.~(\ref{decoupled_space}) yields to the set of equations
\begin{subequations}
	\begin{equation}
	\mathcal{O}(\epsilon^1):\hspace{0.5cm}\mathcal{L}_0 u_1=0,	
	\end{equation}
	\begin{equation}
	\mathcal{O}(\epsilon^2):\hspace{0.5cm}\mathcal{L}_0 u_2+(\mathcal{L}_1+\mathcal{N}_1)u_1=0.	
	\end{equation}	 
\end{subequations}
From the equation at $\mathcal{O}(\epsilon^1)$ one gets that the non-trivial solutions must be proportional to $A_1$, and therefore we can write
\begin{equation}
u_1=A_1 a(X)=\sqrt{\delta}a(X).
\end{equation}
We look for pulse solutions bi-asymptotic to the top homogeneous branch $A^+$, and then we choose $A_1=\sqrt{\delta}$, instead of $A_1=-\sqrt{\delta}$ which corresponds to $A^-.$

Finally, at $\mathcal{O}(\epsilon^2)$ the solvability condition imposes
\begin{equation}
(\mathcal{L}_1+\mathcal{N}_1)u_1=0,
\end{equation}
where
\begin{equation}
\mathcal{L}_1+\mathcal{N}_1=\frac{\beta\sqrt{\delta}}{2}\left[\frac{2D-\beta}{\sqrt{\delta}\beta}\partial_{X}^2-2-a(X)\right],	
\end{equation}
what finally leads to the amplitude equation
\begin{equation}
c_1\partial_X^2 a(X)+c_2 a(X)+c_3 a(X)^2=0,
\end{equation}
which has the same form than Eq.~(\ref{Ampli_trans}), where
\begin{equation}
c_1=\frac{2D-\beta}{\sqrt{\delta}\beta},\hspace{0.4cm}	c_2=-2,\hspace{0.4cm}c_3=-1.
\end{equation}

The pulse solutions
\begin{equation}
a(X)=-\frac{3}{2}\frac{c_2}{c_3}{\rm sech}^2\left(\frac{1}{2}\sqrt{\frac{-c_2}{c_1}}X\right),
\end{equation}
yields, in this case, to 
\begin{equation}
a(x)=-3{\rm sech}^2\left(\frac{1}{2}\sqrt{\frac{2\beta}{2D-\beta}}(\alpha-\alpha_t)^{1/4}x\right),
\end{equation}
which exists always that $2D-\beta>0$.

Hence, a weakly non-linear gap solution of the form 
\begin{equation}\label{small_gap}
A(x)=\frac{\beta}{2}+\sqrt{\alpha-\alpha_t}\left[1+a(x)\right]
\end{equation}
arises from the fold $\alpha_t$ if $2D-\beta>0$. 

\subsection{WEAKLY NON-LINEAR STATES NEAR THE TURING BIFURCATION}\label{WNL_Turing}
Considering that for a fixed value of $\beta$ and $D$ the Turing bifurcation occurs at a given point $(\alpha,A_s)=(\alpha_c,A_c)$, an appropriate asymptotic expansion in term of $\epsilon$ for the different variable reads
	\begin{subequations}
		\begin{equation}\label{expand_Turing_para}
		\alpha=\alpha_c+\delta \epsilon^2,	
		\end{equation}
		\begin{equation}\label{expand_Turing_hom}
		A_s=A_c+\epsilon^2 A_2+\cdots,
		\end{equation}
		\begin{equation}\label{expand_Turing_space}
		u(x,X)=\epsilon u_1(x,X)+\epsilon^2 u_2(x,X)+\epsilon^3 u_3(x,X)+\cdots,	
		\end{equation}
	\end{subequations}
where the space dependent variables are functions of the long scale $X\equiv\epsilon x$ and $x$.

\subsubsection*{\bf Solution of the uniform problem}
Considering the asymptotic expansion (\ref{expand_Turing_hom}) 
we derive the following hierarchical equations for the homogeneous state
\begin{subequations}
	\begin{equation}
	\mathcal{O}(\epsilon^0):\hspace{0.5cm} (-A_c^2+\beta A_c+\alpha_c) A_c=0,	
	\end{equation}
	
	\begin{equation}
	\mathcal{O}(\epsilon^2):\hspace{0.5cm} \mathcal{M}_0 A_2+\delta A_c=0,	
	\end{equation}
	with 
	\begin{equation}
	\mathcal{M}_0=-3A_c^2+2\beta A_c+\alpha_c.
	\end{equation}
\end{subequations}
The solutions at $\mathcal{O}(\epsilon^0)$ gives
\begin{equation}\label{alpha_t_a}
\alpha_c=A_c^2-\beta A_c,
\end{equation}
and $A_c=0$.

The equation at $\mathcal{O}(\epsilon^2)$, leads to the solution 
\begin{equation}
A_2=-\delta\tilde{A}_2=\frac{-\delta A_c}{\mathcal{M}_0}.
\end{equation}

\subsubsection*{\bf Solution of the space dependent problem}
The expansion (\ref{expand_Turing_space}) for space-dependent state implies  $\mathcal{L}=\mathcal{L}_0+\epsilon\mathcal{L}_1+\epsilon^2\mathcal{L}_2+\cdots,$ with 
\begin{subequations}
	\begin{equation}
	\mathcal{L}_0=\mathcal{M}_0+(D-A_c)\partial_x^2-A_c\partial_x^4,
	\end{equation}
	\begin{equation}
	\mathcal{L}_1=2(D-A_c)\partial_X\partial_x-4A_c\partial_X\partial_x^3,
	\end{equation}
	\begin{multline}
	\mathcal{L}_2=(1-2\beta \tilde{A}_2+6A_c\tilde{A}_2)\delta+(D-A_c)\partial_X^2\\-A_2(\partial_x^2+\partial_x^4)-6A_c\partial_x^2\partial_X^2,
	\end{multline}
\end{subequations}
and $\mathcal{N}=\epsilon\mathcal{N}_1+\epsilon^2\mathcal{N}_2+\cdots$ for the non-linear operator,
where
\begin{subequations}
	\begin{equation}
	\mathcal{N}_1=(\beta-3A_c)u_1-u_1(\partial_x^2+\partial_x^4).
	\end{equation}
\begin{multline}
	\mathcal{N}_2=-u_1^2+(\beta-3A_c)u_2-u_2(\partial_x^2+\partial_x^4)\\-2u_1(\partial_X\partial_x+2\partial_X\partial_x^3)
	\end{multline}
\end{subequations}

The insertion of the previous expansion in Eq.~(\ref{decoupled_space}) yields to the set of equations
\begin{subequations}
	\begin{equation}
	\mathcal{O}(\epsilon^1):\hspace{0.5cm}\mathcal{L}_0 u_1=0,	
	\end{equation}
	\begin{equation}\label{Turing_o2}
	\mathcal{O}(\epsilon^2):\hspace{0.5cm}\mathcal{L}_0 u_2+(\mathcal{L}_1+\mathcal{N}_1)u_1=0,	
	\end{equation}
	\begin{equation}\label{Turing_o3}
	\mathcal{O}(\epsilon^3):\hspace{0.5cm}\mathcal{L}_0 u_3+(\mathcal{L}_1+\mathcal{N}_1)u_2+(\mathcal{L}_2+\mathcal{N}_2)u_1=0,	
	\end{equation}		 
\end{subequations}
To solve the $\mathcal{O}(\epsilon^1)$ equation we consider the ansatz:
\begin{equation}
u_1(x,X)=a(X)e^{ik_cx}+c.c.,
\end{equation}
from where we can derive the solvability condition
\begin{equation}
A_c k_c^4+(D-A_c)k_c^2-2\beta A_c-\alpha_c+3A_c^2=0,
\end{equation}
as it was already derived in Sec.~\ref{sec:3}.

At $\mathcal{O}(\epsilon^2)$ the solvability condition is obtained by projecting on the subspace defined by the null eigenvector of the self-adjoint operator $\mathcal{L}_0=\mathcal{L}^\dagger_0$: $w=e^{ik_cx}+c.c.$ To do so we define the standard scalar product in a finite domain with periodic boundary conditions $$\langle f|g\rangle=\frac{1}{L}\int_{-L/2}^{L/2}\bar f(x)\cdot g(x)dx.$$

To calculate this condition first we write
\begin{multline}
(\mathcal{L}_1+\mathcal{N}_1)u_1=f_0|a|^2+f_1i\partial_Xae^{ik_cx}+f_2a^2 e^{2ik_cx}\\+c.c.,
\end{multline}
with 
\begin{subequations}
	\begin{equation}
f_0=2(\beta-3A_c+k_c^2-k_c^4),
	\end{equation}
	\begin{equation}
	f_1=2k_c(D+2k_c^2A_c-A_c),
	\end{equation}
	\begin{equation}
	f_2=f_0/2.
	\end{equation}
\end{subequations}

The solvability condition then implies 
\begin{equation}\label{solv_cond}
\langle w|(\mathcal{L}_1+\mathcal{N}_1)u_1\rangle=0,
\end{equation}
what leads to $f_1=0$, or equivalently to the non-trivial critical wavenumber 
\begin{equation}
k_c^2=\frac{A_c-D}{2A_c}.
\end{equation}

Once this condition is satisfied we can solve Eq.~(\ref{Turing_o2}) considering 
the ansatz
\begin{multline}
u_2(x,X)=W_0|a|^2+W_1ia_Xe^{ik_cx}\\+W_2a^2e^{2ik_cx}+c.c.,
\end{multline}
and matching the coefficients with the same element of the base $\{(e^{ink_cx})\}$, we obtain: 
\begin{equation}
W_0=-f_0/\mathcal{M}_0,
\end{equation}
\begin{equation}
W_1=\frac{-f_1}{\mathcal{M}_0+(A_c-D)k_c^2-A_ck_c^4}=0,
\end{equation}
which follows from the solvability condition (\ref{solv_cond}), and
\begin{equation}
W_2=\frac{-f_2}{\mathcal{M}_0+4(A_c-D)k_c^2-16A_ck_c^4}.
\end{equation}

Finally, in what follows we show how the solvability condition at $\mathcal{O}(\epsilon^3)$ leads to an equation describing the amplitude of the Turing mode $a(X)$.

First, the second term in Eq.~(\ref{Turing_o3}) becomes
\begin{equation}
(\mathcal{L}_1+\mathcal{N}_1)u_2=g_0+g_1e^{ik_cx}+g_2e^{2ik_cx}+g_3e^{3ik_cx},
\end{equation}
with 
\begin{multline}
g_1(X)=g_1^c|a(X)|^2a(X)=\\\left[(\beta-3A_c)(W_0+W_2)+(4k_c^2-16k_c^4)W_2\right]|a(X)|^2a(X),
\end{multline}
whereas
the third term becomes
\begin{equation}
(\mathcal{L}_2+\mathcal{N}_2)u_1=h_0+h_1e^{ik_cx}+h_2e^{2ik_cx}+h_3e^{3ik_cx},
\end{equation}
with 
\begin{equation}
h_1=h_1^a\delta a(X)+h_1^b\partial_X^2a(X)+h_1^c|a(X)|^2a(X),
\end{equation}
and
\begin{subequations}
	\begin{equation}
	h_1^a=(1-2\beta\tilde{A}_2+6A_c\tilde{A}_2-k_c^2\tilde{A}_2+k_c^4\tilde{A}_2)
	\end{equation}
	\begin{equation}
	h_1^b=(D-A_c+6A_ck_c^2),
	\end{equation}
	\begin{equation}
	h_1^c=(\beta-3A_c)(W_0+W_2)+(k_c^2-k_c^4)(W_0+W_2)-3
	\end{equation}
\end{subequations}
  \begin{figure}[!t]
  	\centering
  	\includegraphics[scale=1]{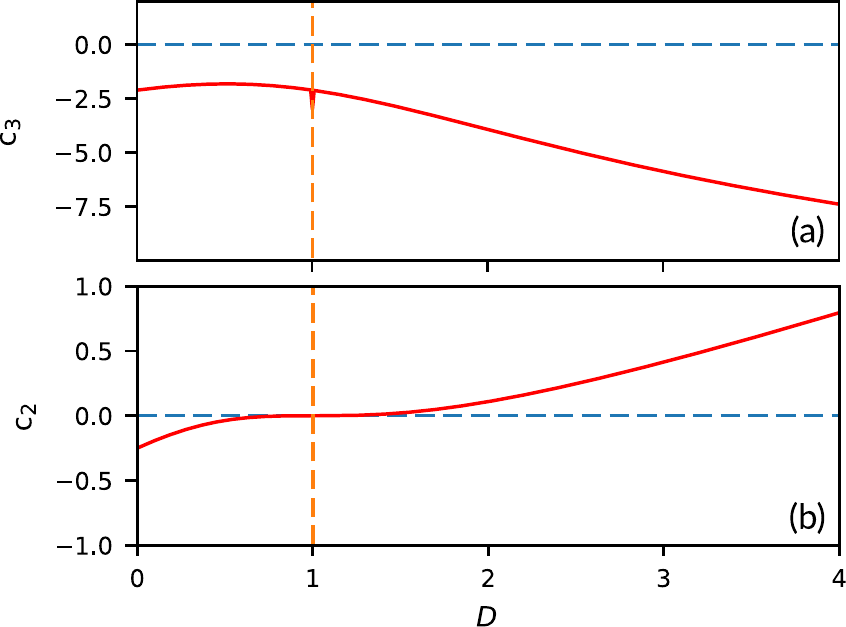}
  	\caption{(Color online) Coefficients $c_3$ in (a) and $c_2$ in (b) as a function of $D$ for $\beta=2$. The vertical orange dashed line represents the condition $D=\beta/2$. Thus $c_2$ is negative for $D<\beta/2$ and positive otherwise. For the range of parameter considered in this work $c_3$ is always negative. }
  	\label{WN}
  \end{figure}
The solvability condition 
\begin{multline}
\langle w|(\mathcal{L}_1+\mathcal{N}_1)u_2\rangle+\langle w|(\mathcal{L}_2+\mathcal{N}_2)u_1\rangle=0,
\end{multline}
then yields to the amplitude equation 
\begin{equation}\label{amplitude_Turing}
\delta a(X)+c_2\partial_X^2 a(X)+c_3 |a(X)|^2a(X)=0,
\end{equation}
with 
\begin{subequations}
	\begin{equation}
	c_2=h_1^b/h_1^a,
	\end{equation}
	\begin{equation}
	c_3=(g_1^c+h_1^c)/h_1^a.
	\end{equation}
\end{subequations}

Due to the complex form of (\ref{expAC}) the simplification of the coefficients $c_3$ and $c_2$ to a simple expression of $\beta$ and $D$ is not possible.
However, using symbolic software we find that $c_2$ cancels out exactly at $D=D_q\equiv\beta/2$ [see Sec.~\ref{sec:2}], and is negative ($c_2<0$) for $D<\beta/2$, and positive otherwise.

We can solve Eq.~(\ref{amplitude_Turing}) by considering the ansatz $a(X)=\psi(X) e^{i\varphi}$. 
If $\psi\neq\psi(X)$ then the amplitude equation reduces to 
\begin{equation}
\psi(\delta+c_3\psi^2)=0,
\end{equation}
what implies the solutions $\psi=0$ and $\psi=\sqrt{-\delta/c_3}$.
This solution corresponds to a periodic gap pattern of the form
\begin{equation}\label{pattern}
A(x)=A_c+(\alpha-\alpha_c)\tilde{A}_2+2\sqrt{\frac{\alpha-\alpha_c}{-c_3}}{\rm cos}(k_c x+\varphi),
\end{equation}
that arises sub- or supercritical depending on the sign of the coefficient $c_3$: if $c_3<0$ the pattern arises subcritically and supercritically otherwise. Figure~\ref{WN}(a) shows $c_3$ as a function of $D$ for $\beta=2$. As we can observe, this coefficient is negative for any value of $D$, what means that the periodic pattern is born subcritically from the TI. For the range of $\beta$ studied in this manuscript $c_3$ is always negative.

In the subcritical regime, moreover, Eq.~(\ref{amplitude_Turing}) has also a solution of the form
\begin{equation}
\psi(X)=\sqrt{\frac{-2\delta}{c_3}}{\rm sech}\left(\sqrt{\frac{-\delta}{c_2}}X\right),
\end{equation}
which corresponds to the gap state 
\begin{multline}\label{gap}
A(x)=A_c+(\alpha-\alpha_c)\tilde{A}_2+\\2\sqrt{\frac{2(\alpha-\alpha_c)}{-c_3}}{\rm sech}\left(\sqrt{\frac{\alpha-\alpha_c}{-c_2}}x\right){\rm cos}(k_cx+\varphi).
\end{multline}

This state arises subcritically from the TI together with the gap pattern (\ref{pattern}), and exists whenever $c_2$ is negative. In Fig.~\ref{WN}(b) we plot $c_2$ as a function of $D$ for $\beta=2$. $c_2$ cancels out at exactly $D=\beta/2$, as is marked with vertical orange dashed line. Therefore gap states of the form (\ref{gap}) exist if $D<\beta/2$. 



\bibliographystyle{ieeetr}
\bibliography{dryland_v2}
\end{document}